\documentclass[pra,showpacs,superscriptaddress,twocolumn]{revtex4}

\usepackage{dcolumn,graphicx}

\begin{document}

\begin{widetext}

\title{Quantum Cryptography: Security Criteria Reexamined}

\author{Dagomir Kaszlikowski}
\affiliation{Department of Physics, National University of Singapore,
Singapore 117\,542, Singapore}

\author{Ajay Gopinathan}
\affiliation{Department of Physics, National University of Singapore,
Singapore 117\,542, Singapore}
\affiliation{National Institute of Education,
Nanyang Technological University, Singapore 637616, Singapore}

\author{Yeong Cherng Liang}
\affiliation{Department of Physics, National University of Singapore,
Singapore 117\,542, Singapore}

\author{L.~C. Kwek}
\affiliation{National Institute of Education,
Nanyang Technological University, Singapore 637616, Singapore}
\affiliation{Department of Physics, National University of Singapore,
Singapore 117\,542, Singapore}

\author{Berthold-Georg~Englert}
\affiliation{Department of Physics, National University of Singapore,
Singapore 117\,542, Singapore}

\date{27 May 2004}

\begin{abstract}
We find that the generally accepted security criteria are flawed for a whole
class of protocols for quantum cryptography.
This is so because a standard assumption of the security analysis, 
namely that the so-called square-root measurement is optimal for eavesdropping 
purposes, is not true in general. 
There are  rather large parameter regimes in which the optimal measurement
extracts substantially more information than the square-root measurement. 
\end{abstract}


\maketitle

\end{widetext}

\section{Introduction}\label{sec:intro}
All practical implementations of protocols for quantum cryptography have to
deal with the unavoidable noise in the transmission lines, and possibly the
intervention of an eavesdropper, that degrade the correlations in the raw-key 
data of the communicating parties --- Alice and Bob. 
They then face a double task: 
First, they must establish how much Eve, the evildoing eavesdropper, can
possibly know about their data; and second, they must extract a secure
noise-free key sequence from the insecure noisy raw data.

The second task of key generation is solved by exploiting the findings and
methods of classical information theory, in particular the lesson of the
seminal work by Csisz\'ar and K\"orner \cite{CK}.
They demonstrated that Alice and Bob can always generate a secure key,
provided that the mutual information between them exceeds the mutual
information between either one of them and Eve.

The first task of determining how much Eve knows thus amounts to figuring out
the maximally attainable mutual information between her and either Alice or
Bob.
There are two different, but equivalent, lines of reasoning that one can
choose to follow, depending on how one pictures the communication between
Alice and Bob, and Eve's tampering with it.

One scenario is that of the 1984 protocol by Bennett and Brassard (BB84, 
\cite{BB84}), in which Alice sends quantum-information carriers to Bob
through an appropriate, authenticated quantum channel.
Eve intercepts each carrier in transmission and keeps an imperfect copy,
obtained by operating a quantum-cloning machine, before forwarding the carrier
to Bob.
The quest is then for the best cloning machine 
--- best for \emph{this} purpose ---
in conjunction with the best way of extracting information from the clones.

The other scenario is that of the 1991 protocol by Ekert (E91,
\cite{Ekert91}), in 
which a source distributes entangled pairs of carriers to Alice and Bob, who
make statistically independent measurements on them, thereby effectively
establishing a quantum channel between themselves.
Eve is given full control of the source.
She keeps a quantum record of what is sent in the form of auxiliary quantum
systems, usually termed \emph{ancillas}, that she entangles with the paired
carriers. 
Here the quest is for the best ancilla states in conjunction with the best way
of extracting information from the ancillas. 

In lack of superior alternatives, 
the standard analysis of protocols of BB84 type
invokes unproven assumptions about optimal cloning machines; see, for
example, Refs.~\cite{Bourennane+4:02,Cerf+3:02} and the recent
paper by Ac\'\i{}n \textit{et al.} \cite{Acin+2:03}.
Likewise, there is a common assumption in the analysis of E91-type protocols,
namely that the so-called square-root measurement (SRM, \cite{Chefles:00}) 
is optimal for Eve's processing of the ancillas; see the recent paper by Liang
\textit{et al.} \cite{Liang+4:03}, for example.
The established equivalence of the BB84 and E91 scenarios \cite{E91=BB84}, 
and the fully equivalent security criteria thus found, is strong circumstantial
evidence that these assumptions ---  
about Eve's best intercept strategy and her best way of processing the
ancillas, respectively 
--- are equivalent as well.

It is the objective of this article to demonstrate that the SRM is \emph{not}
optimal for a whole class of quantum cryptography protocols, the tomographic
protocols of Refs.\ \cite{Liang+4:03,Bruss+5:03}; it may very well not be
optimal for other protocols, too. 
The equivalence stated above then implies the well-founded conjecture that 
there are also better intercept strategies than those usually regarded as
best.
We offer some remarks about the connection of this work with intercept
strategies in the Appendix.

\section{The pyramid of ancilla states}\label{sec:pyramid}
\begin{figure}[t]
\centering\includegraphics{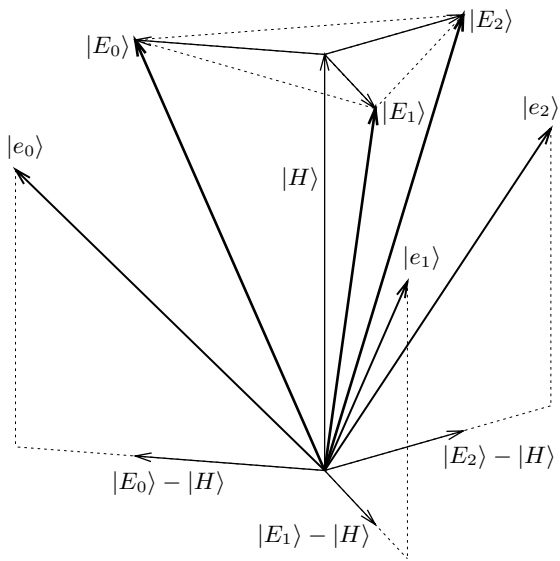}  
\caption{\label{fig:trine}%
Pyramid geometry for $N=3$.
The ancilla kets $\ket{E_k}$, of unit length, 
are the edges of the ancilla pyramid.
Its shape is determined by the parameter $\lambda$ of \eqref{1}, 
the cosine of the acute angle between any pair of edges.
The height ket $\ket{H}$ of \eqref{2} points from the tip of the pyramid to
the center of its base; its length is $\sqrt{r_0}$. 
The kets $\ket{E_k}-\ket{H}$, of length $\sqrt{1-r_0}$, point from the center 
of the pyramid base to its corners. 
The SRM kets $\ket{e_k}$ of \eqref{8a}, of unit length, 
define the SRM pyramid, which has right angles between its edge kets.
The SRM pyramid is wider than, but not as high as, the ancilla pyramid. 
}
\end{figure}

We build on the work of Ref.\ \cite{Liang+4:03}, 
where the protocols are phrased as 
generalizations of the E91 scenario to $N$ letter alphabets ($N=2,3,\dots$), 
The source controlled by Eve would emit pairs of qubits for $N=2$, 
pairs of qutrits for $N=3$, \dots, pairs of \emph{qunits} in the general case.
After everything is done and said,
Eve knows that her ancilla is in the state described by ket $\ket{E_k}$ 
if Alice obtains value $k$ for her qunit of the respective pair 
(with $k=0,1,\dots,N-1$).
Since there is a common (real) angle between every pair of ancilla states,
\begin{eqnarray}
  \label{eq:1}
  \braket{E_k}{E_l}&=&\lambda+(1-\lambda)\delta_{kl}=\left\{
    \begin{array}{c@{\mbox{\ if\ }}c}
1 & k=l\\ \lambda & k\neq l
      \end{array}\right\}\nonumber\\
&=&r_0-r_1+Nr_1\delta_{kl}\,,
\end{eqnarray}
the $N$ ancilla kets can be regarded as the edges of an $N$-dimensional pyramid
\cite{notation}; 
see Fig.~\ref{fig:trine} for an illustration of the case of $N=3$. 
The average ancilla ket
\begin{equation}
  \label{eq:2}
  \ket{H}=\frac{1}{N}\sum_{k=0}^{N-1}\ket{E_k}
\end{equation}
points from the tip of the pyramid to the center of its $(N-1)$-dimensional
base \cite{base}, so that the length of $\ket{H}$,
$\sqrt{\braket{H}{H}}=\sqrt{r_0}\,$, is the height of the pyramid. 
The pyramid volume is given by
$(1/N!)(Nr_0)^{1/2}(Nr_1)^{(N-1)/2}$, it is largest for $\lambda=0$, 
$r_0=r_1=1/N$ when the pyramid is a corner of a $N$-dimensional cube.

Geometry restricts $\lambda$ to the range $-1/(N-1)\leq\lambda\leq1$,
where both limits correspond to degenerate pyramids that have no
$N$-dimensional volume. 
For $\lambda=1$, we have a single ancilla state and the pyramid is just a
line, a pyramid of unit height and no base; 
and for $\lambda=-1/(N-1)$ we have linearly dependent ancilla kets that
span an $(N-1)$-dimensional subspace, so that the pyramid has no height.
In the context of quantum cryptography, however, only nonnegative $\lambda$ 
values are relevant, for which $r_0\geq r_1$.
In other words, the pyramids of interest are acute, in the sense that 
the  common  angle between each pair of their edges is acute. 

Alice gets each $k$ value with probability $1/N$, so that 
\begin{equation}
  \label{eq:3}
  \rho=\frac{1}{N}\sum_{k=0}^{N-1}\ket{E_k}\bra{E_k}
\end{equation}
is the statistical operator for Eve's ancillas.
The height ket $\ket{H}$ of \eqref{2} is eigenket of $\rho$ to eigenvalue
$r_0$ and all kets orthogonal to $\ket{H}$ are eigenkets to the $(N-1)$-fold
degenerate eigenvalue $r_1=r_0-\lambda=(1-\lambda)/N$.

The $N$ kets $\ket{E_k}-\ket{H}$, 
each of length $\sqrt{1-r_0}=\sqrt{(N-1)r_1}$, 
point from the center of the ancilla-pyramid base to its corners. 
They span the $(N-1)$-dimensional subspace to eigenvalue $r_1$.

\section{Which edge of the pyramid?}\label{sec:edge}
\subsection{The pretty good square-root measurement}\label{sec:SRM}
Eve extracts information out of $\rho$ with the aid of a generalized 
measurement, a positive-operator-valued measure (POVM), specified 
by a decomposition of the identity in the $N$-dimensional ancilla space 
into $M$ nonnegative operators,
\begin{equation}
  \label{eq:4}
  1=\sum_{m=0}^{M-1}P_m\,,\quad P_m\geq0\,.
\end{equation}
The mutual information between Alice and Eve,
\begin{equation}
  \label{eq:5}
  I=\sum_{n=0}^{N-1}\sum_{m=0}^{M-1}p_{nm}
\log_N\frac{p_{nm}}{p_{n\cdot}p_{\cdot m}}\,,
\end{equation}
is then computable from the joint probabilities
\begin{equation}
  \label{eq:6}
  p_{nm}=\frac{1}{N}\bra{E_n}P_m\ket{E_n}
\end{equation}
and their marginals
\begin{equation}
  \label{eq:7}
  p_{n\cdot}=\sum_{m=0}^{M-1}p_{nm}=\frac{1}{N}\,,\quad  
  p_{\cdot m}=\sum_{n=0}^{N-1}p_{nm}\,.
\end{equation}
For convenient normalization, the logarithm in \eqref{5} is taken to base $N$,
so that $I\leq1$ with the maximum achieved for uniform perfect correlations,
that is for $M=N$ and $p_{nm}=\delta_{nm}/N$. 

\begin{figure}[t]
\centering\includegraphics{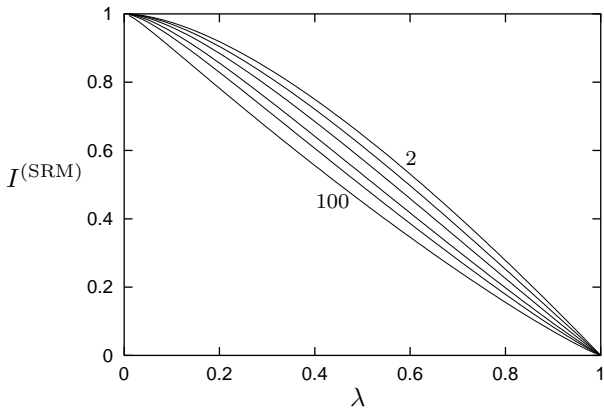}  
\caption{\label{fig:SRM}%
Mutual information between Alice and Eve if Eve performs the square-root
measurement. 
The curves refer to $N=2,3,5,10,20$, and $100$, and the plot covers the
range $0\leq\lambda\leq1$ that is relevant for quantum cryptography.}
\end{figure}

The POVM for the SRM is specified by setting $M=N$ and
\begin{equation}
  \label{eq:8}
  P_m=(N\rho)^{-1/2}\ket{E_m}\bra{E_m}(N\rho)^{-1/2}\equiv\ket{e_m}\bra{e_m}
\end{equation}
with
\begin{equation}
    \label{eq:8a}
\ket{e_m}=\Bigl(\ket{E_m}-\ket{H}\Bigr)\frac{1}{\sqrt{Nr_1}}
+\ket{H}\frac{1}{\sqrt{Nr_0}}
\,.
\end{equation}
The resulting joint probabilities are
\begin{equation}
  \label{eq:9}
  p_{nm}=\frac{1}{N}\left|\braket{E_n}{e_m}\right|^2
=\frac{1}{N}\bigl[\eta_1+(\eta_0-\eta_1)\delta_{nm}\bigr]\,,
\end{equation}
where
\begin{equation}
  \label{eq:10}
\sqrt{\eta_0}-\sqrt{\eta_1}=\sqrt{Nr_1}\quad\mbox{and}\enskip
\eta_0+(N-1)\eta_1=1\,.
\end{equation}
We note that the SRM thus associated with the ancilla pyramid 
happens to be a standard von Neumann measurement, not a POVM proper, 
because the projectors in \eqref{8} are pairwise orthogonal, 
$\mathrm{tr}\left\{P_mP_{m'}\right\}=\delta_{mm'}$.
The mutual information acquired by performing the SRM,
\begin{equation}
  \label{eq:11}
 I^\mathrm{(SRM)}=\eta_0\log_N(N\eta_0)+(N-1)\eta_1\log_N(N\eta_1)\,, 
\end{equation}
is shown in Fig.~\ref{fig:SRM} for $N=2,3,5,10,20,100$.

\subsection{Better than pretty good}\label{sec:opt}
Whereas the SRM is known to be ``pretty good'' as a rule \cite{PrettyGood},
it is also known that it does not always optimize the mutual information.
In particular, Shor has pointed out that there are superior POVMs for $N=3$
and some $\lambda<0$, and has conjectured that there is also a $\lambda>0$
range in which other POVMs could be better \cite{Shor:02}.
Shor's explicit example for $\lambda<0$ is interesting in its own right but
does not seem to have any bearing on the security analysis of 
quantum-cryptography protocols.
By contrast, the $\lambda>0$ examples reported below, are of immediate
relevance, as they invalidate, at least partly, established security criteria.

Consider the one-parametric family of POVMs defined by $M=N+1$ and
$P_m=\ket{\bar{e}_m}\bra{\bar{e}_m}$ with
\begin{eqnarray}
  \label{eq:12}
m<N:&&  
\ket{\bar{e}_m}=\Bigl(\ket{E_m}-\ket{H}\Bigr)\frac{1}{\sqrt{Nr_1}}
+\ket{H}\frac{t}{\sqrt{Nr_0}}\,,\nonumber\\
m=N: && \ket{\bar{e}_N}=\ket{H}\sqrt{\frac{1-t^2}{r_0}}\,,
\end{eqnarray}
where $0\leq t\leq1$. 
The SRM kets of \eqref{8a} obtain for $t=1$. 

For $t<1$, the measurement pyramid, which has the kets $\ket{\bar{e}_0}$,
\dots,$\ket{\bar{e}_{N-1}}$ for its edges,
has the same base area as the SRM pyramid, but is of smaller height 
and therefore obtuse.
Since the angle between any such given $\ket{\bar{e}_m}$ and the ancilla kets
$\ket{E_n}$ with $n\neq m$ increases as $t$ decreases from $t=1$, the sector
of $m<N$ will have increased mutual information. 
But this comes at a price: When Eve finds $\ket{\bar{e}_N}\propto\ket{H}$ 
she has no clue about Alice's value; the sector $m=N$ 
is inconclusive and provides no contribution at all to the mutual information. 
Accordingly, the optimal choice of $t$ is such that the increase of mutual
information in the $m<N$ sector is balanced against the increase in the 
probability of the inconclusive result; this probability equals $(1-t^2)r_0$.

\begin{figure}[t]
\centering\includegraphics{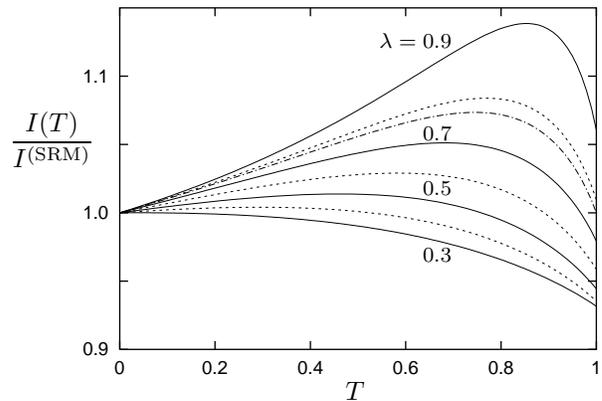}  
\caption{\label{fig:InfoOfT}%
Mutual information for the POVM of \eqref{12} relative to that of the SRM. 
For $N=10$, the plot shows the ratio of $I(T)/I^\mathrm{(SRM)}$
as a function of $T$ for $\lambda=0.9,0.7,0.5,0.3$ (solid lines) and
for  $\lambda=0.8,0.6,0.4$ (dashed lines). 
The left end ($T=0$) refers to the SRM, the right end ($T=1$) to the MUD.
For $\lambda=0.77276$ (dash-dotted line), 
both give the same mutual information.
}
\end{figure}

For $t=\sqrt{r_1/r_0}$, the POVM specified by \eqref{12} is the 
``measurement for unambiguous discrimination'' (MUD, \cite{Chefles:00}), 
for which $\braket{E_n}{\bar{e}_m}=0$ if $n\neq m<N$, so that there are perfect
correlations, and thus full mutual information, in the $m<N$ sector.
The cost for this perfection is, however, so high that the MUD never maximizes
the mutual information, although it can outperform the SRM. 
The optimal choice for $t$ is always in the range $\sqrt{r_1/r_0}<t\leq1$.
This observation is illustrated in Fig.~\ref{fig:InfoOfT} for $N=10$ and
various values of $\lambda$, including $\lambda=0.77276$, for which the MUD
and the SRM give the same mutual information.
The plot shows only the $t$ range of interest, conveniently re-parameterized
in terms of $T$, a scaled version of $t$, introduced in accordance with
\begin{equation}
  \label{eq:13}
  t=1-T+T\sqrt{r_1/r_0}\,.
\end{equation}
Thus, $T=0$ refers to the SRM, and $T=1$ to the MUD.
 
The mutual information for the POVMs specified by \eqref{12} is given by
\begin{eqnarray}
  \label{eq:14}
  I(T)&=&\bar{\eta}_0\log_N\frac{N\bar{\eta}_0}{\bar{\eta}_0+(N-1)\bar{\eta}_1}
\nonumber\\&&
+(N-1)\bar{\eta}_1\log_N\frac{N\bar{\eta}_1}{\bar{\eta}_0+(N-1)\bar{\eta}_1}\,,
\end{eqnarray}
where
\begin{equation}
  \label{eq:15}
  \bar{\eta}_0= \bigl(\sqrt{\eta_0}-T\sqrt{\eta_1}\bigr)^2\,,\quad
  \bar{\eta}_1= (1-T)^2\eta_1
\end{equation}
are the $T$ dependent versions of $\eta_0$, $\eta_1$.
For ancilla pyramids with a large volume, 
$0<\lambda<(3-4/N)/(N-1)\equiv\Lambda$,
the maximum of $I(T)$ obtains for $T=0$, which is to say that the SRM is
optimal in this range of small $\lambda$ values.
By contrast, for ancilla pyramids with a rather small volume, 
$\Lambda<\lambda<1$, the maximum of $I(T)$ is reached for 
$T=1-(\sqrt{\eta_0/\eta_1}-1)/(N-2)$, that is when the arguments of the two
logarithms in \eqref{14} equal $N-1$ and  $1/(N-1)$, respectively.
Then, the measurement pyramid is obtuse.

In summary we have
\begin{eqnarray}
  \label{eq:16}
  I_\mathrm{max}&\equiv&\max_TI(T)\\\nonumber&=&\left\{
    \begin{array}{l}
\mbox{$I^\mathrm{(SRM)}$ of \eqref{11} 
if $\displaystyle0\leq\lambda\leq\Lambda=\frac{3N-4}{N(N-1)}$},\\[1ex]
\mbox{$\displaystyle(1-\lambda)\frac{N-1}{N-2}\log_N(N-1)$
if $\Lambda\leq\lambda\leq1$}.
    \end{array}\right.
\end{eqnarray}
This is our central result.

\begin{figure}[t]
\centering\includegraphics{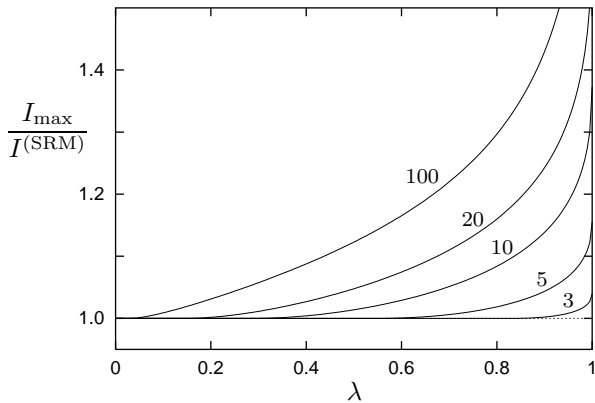}  
\caption{\label{fig:ratio}%
Ratio of the maximal mutual information $I_\mathrm{max}$ and the SRM value 
$I^\mathrm{(SRM)}$, for $N=3,5,10,20,100$, as a function of $\lambda$.
}
\end{figure}

For $\lambda$ values that exceed the threshold value of $\Lambda$
substantially, the optimal POVM from the family \eqref{12}
gives significantly more mutual information than the SRM.
This can be seen by plotting the ratio $I_\mathrm{max}/I^\mathrm{(SRM)}$ 
as a function of $\lambda$; see Fig.~\ref{fig:ratio}.
The $\lambda\to1$ limit,
\begin{equation}
  \label{eq:17}
  \frac{I_\mathrm{max}}{I^\mathrm{(SRM)}}\to\frac{N/2}{N-2}\ln(N-1)
\quad\mbox{as $\lambda\to1$},
\end{equation}
shows that the optimal POVM provides much more information than the SRM if $N$
is large, and then the range $0\leq\lambda<\Lambda\simeq3/N$ is small 
in addition.

\section{Summary and discussion}\label{sec:summary}
In summary, there are POVMs that outperform the SRM for $\lambda>\Lambda$, 
and we know the optimal POVM of the sort defined by \eqref{12} quite 
explicitly.
We are, in fact, quite sure that it is the global optimum because an
extensive numerical search failed to find any better POVM.

A first search covered a large class of POVMs that respect the geometry of 
the ancilla pyramid:
We took parameter $t$ to be complex; we rotated around the symmetry axis
specified by ket $\ket{H}$; and we considered weighted sums of several such
POVMs, with different $t$ parameters and different rotations.
For all of the many $N$ and $\lambda$ values, for which 
the numerical investigation was performed, the
optimal POVM was always of the kind described above.

A second search, not restricted by geometrical or other constraints, confirmed
these findings. 
It used the numerical method of Ref.~\cite{iteration}, which is a
fix-point iteration that converges monotonically toward the optimal POVM.

We note further that the large relative difference shown in
Fig.~\ref{fig:ratio} occurs where both $I_\mathrm{max}$ and $I^\mathrm{(SRM)}$
are small, and so the absolute difference is rather small 
(see the figure in Ref.~\cite{announce}).
Therefore, the SRM threshold values given in Table~I 
of Ref.~\cite{Liang+4:03} are quite good approximations for the true
threshold values, as shown by the numerical values in 
Table~\ref{tbl:newthresholds}.

\begin{table}[b]
\begin{ruledtabular}
\caption{\label{tbl:newthresholds}%
Threshold values for the disturbance below which the 
Csisz\'ar--K\"orner theorem ensures that a secure key can be 
extracted from the noisy raw data.
The second column gives the critical disturbance, that is
$(N-2)^2/[(N-2)^2+N]$, above which the SRM is optimal, 
as implied by Eq.~\eqref{16}.
The third column repeats the values of Refs.~\cite{Cerf+3:02} and
\cite{Liang+4:03}, where Eve extracts information with the aid of the SRM.
The true threshold values of the fourth column obtain for the optimal POVM. }
\begin{tabular}{@{\qquad}rddd@{\qquad}}
& \multicolumn{1}{c}{Critical} &
\multicolumn{2}{c}{Csisz\'ar--K\"orner thresholds}\\
$N$ & \multicolumn{1}{c}{value} & \multicolumn{1}{r}{SRM\phantom{0\%}} 
& \multicolumn{1}{r}{true\qquad\phantom{0\%}}\\
\hline
  2 &  0.0\% & 15.6373\% & 15.6373\% \\ 
  3 & 25.0\% & 22.6714\% & 22.6707\% \\ 
  4 & 50.0\% & 26.6561\% & 26.5989\% \\ 
  5 & 64.3\% & 29.2303\% & 29.1038\% \\ 
 10 & 86.5\% & 34.9713\% & 34.7051\% \\ 
 30 & 96.3\% & 39.8403\% & 39.6259\% \\ 
 50 & 97.9\% & 41.1886\% & 41.0284\% \\ 
100 & 99.0\% & 42.5282\% & 42.4295\% \\
$\infty$ & 100.0\% & 50.0000\% & 50.0000\%
\end{tabular}
\end{ruledtabular}
\end{table}

The ``disturbance'' values listed in this table are the quantities denoted by
$D^\mathrm{ind}_{d+1}$ in Ref.~\cite{Cerf+3:02} and by $1-\beta_0$ in
Ref.~\cite{Liang+4:03}, respectively.
There is no difference for $N=2$, of course, but for all $N>2$ the true
threshold is noticeably lower than the SRM threshold.
In addition to this shift of the threshold, there is also a reduced efficiency 
inside the Csisz\'ar-K\"orner regime (below the threshold) and this must be
taken into account when extracting the secure key sequence from the noisy raw
data. 
Fortunately, however, almost all of the practical quantum cryptography scheme 
presently implemented use qubits ($N=2$), and then the SRM \emph{is} optimal.
Also, the optimal POVMs have no bearing on the threshold for classical
advantage distillation \cite{Acin+2:03,Bruss+5:03}, because the SRM remains
optimal in the relevant limit, even for coherent eavesdropping 
attacks \cite{ADneqED}.
   
In the spirit of Shor's investigation of obtuse pyramids,
the eavesdropping procedure presented here can be viewed as a quantum
communication channel, in which Alice transmits nonorthogonal and equally
distributed signal states to Eve. 
The amount of information about the sequence of states sent by Alice,
maximized over all possible POVMs, is then the \emph{accessible information}
of this quantum channel. 
Therefore, the maximal mutual information \eqref{16} between Alice and Eve 
gives us also this accessible information for $0\leq \lambda$, which
supplements, for $N=3$, Shor's $\lambda<0$ result.

\begin{acknowledgments}
We wish to thank Antonio Ac\'\i{}n, Thomas Durt, 
and Jaroslav \v{R}eh\'a\v{c}ek for valuable discussions. 
We gratefully acknowledge the financial support from 
A$^*$Star Grant No.\ 012-104-0040 and from
NUS Grant WBS: R-144-000-089-112.
\end{acknowledgments}

\appendix*

\section{Intercept attacks}
Here are a few remarks about the connection with intercept attacks on qunits
sent through an authenticated quantum channel.
We make use of the notational conventions of Ref.~\cite{Liang+4:03} without
explaining them anew, and refer to Eq.~(12), say, of Ref.~\cite{Liang+4:03} by
\aref{12}.

The geometry of the unnormalized ancilla states $\ket{\tilde{E}^{(m)}_{kl}}$
is completely determined, for a given $m$ value, by the inner products of
Eq.~\aref{6}, and Eq.~\aref{7} states the transformation law between ancilla 
states to different $m$ values.
It follows from this equation that the $k$ index of
$\ket{\tilde{E}^{(m)}_{kl}}$ is analogous to that in $\ket{\overline{m}_k}$,
and the $l$ index to that in $\ket{m_l}$.
Therefore, it is expedient to regard the $\ket{\tilde{E}^{(m)}_{kl}}$'s 
as the kets of two-qunit states that are superpositions of basis kets
of the $\ket{\overline{m}_km_l}$ kind.
They then acquire the strikingly simple explicit form
\begin{equation}
  \label{eq:A1}
  \ket{\tilde{E}^{(m)}_{kl}}=\ket{\overline{\psi}}\delta_{kl}\frac{a}{\sqrt{N}}
+\ket{\overline{m}_km_l}\frac{b}{N}\,, 
\end{equation}
where
\begin{equation}
  \label{eq:A2}
  \ket{\overline{\psi}}=\frac{1}{\sqrt{N}}\sum_k\ket{\overline{m}_km_k}
\qquad\text{(any $m$ value)}
\end{equation}
is the maximally entangled state that is conjugate to $\ket{\psi}$ 
of Eq.~\aref{2}.
This ansatz for $\ket{\tilde{E}^{(m)}_{kl}}$ is consistent with Eq.~\aref{6} 
if the complex amplitudes $a,b$ obey
\begin{equation}
  \label{eq:A3}
  \Bigl|a+\frac{1}{N}b\Bigr|^2=\beta_0-\frac{N-1}{N}\beta_1\,,
\qquad \bigl| b \bigr|^2=N\beta_1\,,
\end{equation}
but no other restrictions apply, so that $a=\sqrt{\beta_0-\beta_1}$,
$b=i\sqrt{N\beta_1}$ is a permissible choice.

The entangled pure state $\ket{\Psi}$ of Eq.~\aref{5} that is prepared by Eve 
is then of the compact form
\begin{equation}
  \label{eq:A4}
  \ket{\Psi}=\ket{\psi^{\ }_{12}\overline{\psi}_{34}}a
            +\ket{\psi^{\ }_{13}\overline{\psi}_{24}}b\,,
\end{equation}
where qunit~1 is sent to Alice, qunit~2 is sent to Bob, and qunits~3 and 4
make up Eve's ancilla.
We note that this is the generic form of $\ket{\Psi}$ because \emph{all}
alternatives are obtained from this $\ket{\Psi}$ by unitary transformations on
the ancilla. 

Now, the ``asymmetric universal quantum cloning machines'' \cite{asym-ucm},
generalizations of the symmetric ones introduced by Bu\v{z}ek and Hillery
\cite{sym-ucm}, that are employed in Refs.~\cite{Bourennane+4:02,Cerf+3:02}
for the analysis of intercept attacks on the qunit in transmission from Alice 
to Bob, are characterized by a four-qunit state of the form \eqref{A4}.
The resulting states of the clone-anticlone pair are thus fully analogous to
the ancilla states $\ket{\tilde{E}^{(m)}_{kl}}$ in \eqref{A1}.
Of those, the ones with $k\neq l$ are orthogonal among themselves and
orthogonal to those with $k=l$, and the latter form the pyramid of ancilla
states described in Sec.~\ref{sec:pyramid}.
Accordingly, Eve can extract more information if she applies the optimal POVM
of Sec.~\ref{sec:opt} to the clone-anticlone pair, rather than submitting them
to the usual SRM.


\begin{thebibliography}{99}
\bibitem{notation}
Put $N\to n$, $\ket{E_k}\to\ket{E_{kk}}$, and $\lambda\to1-\beta_1/\beta_0$ to
convert to the notational conventions of Ref.\ \cite{Liang+4:03}.
Other symbols, in particular $r_0$, $r_1$ and $\eta_0$, $\eta_1$, have the same
significance.
\bibitem{base}
Geometrically speaking, the base of the $N$-dimensional pyramid is itself an
$(N-1)$-dimensional pyramid with edges of length $\sqrt{2Nr_1}$ 
and an angle of $60^\circ$ between them.
\end{thebibliography}
\end{document}